\documentclass[useAMS,usenatbib]{mnras}

\usepackage{times}
\usepackage{graphics,epsfig}
\usepackage{graphicx}
\usepackage{amsmath}
\usepackage{amssymb}
\usepackage{bm}
\usepackage{dcolumn}
\usepackage{epsfig}
\usepackage{subfig}
\usepackage{tabularx}
\usepackage[normalem]{ulem}
\usepackage{times}
\usepackage{ulem}
\usepackage{array}

\newcommand{\kms}{km~s$^{-1}$}

\newcommand{\arcs}{\ensuremath{^{\prime\prime}}}

\newcommand{\hii}{H{\sc i} 21\,cm }
\newcommand{\hi}{H{\sc i} }
\newcommand{\cm}{cm$^{-2}$}

\title[Host galaxy of AT2018cow in H{\sc i}]{H{\sc i} 21\,cm mapping of the host galaxy of AT2018cow: a fast-evolving luminous transient within a ring of high column density gas}

\author[Roychowdhury et al.]{Sambit Roychowdhury$^{1}$\thanks{sambit.roychowdhury@ias.u-psud.fr}, Maryam Arabsalmani$^{2,3,4}$, Nissim Kanekar$^{5}$\\
$^1$ Institut d'Astrophysique Spatiale, CNRS, Universite Paris-Sud, Universite Paris-Saclay, Bet. 121, 91405 Orsay Cedex, France \\
$^2$ IRFU, CEA, Universit\'e Paris-Saclay, F-91191 Gif-sur-Yvette, France\\
$^3$ Universit\'e Paris Diderot, AIM, Sorbonne Paris Cit\'e, CEA, CNRS, F-91191 Gif-sur-Yvette, France\\
$^4$ School of Physics, The University of Melbourne, VIC 3010, Australia\\
$^5$ National Centre for Radio Astrophysics, Tata Institute of Fundamental Research, Pune University, Pune 411007, India }

\begin{document}
\date{}

\pagerange{\pageref{firstpage}--\pageref{lastpage}} \pubyear{}

\maketitle

\label{firstpage}

\begin{abstract}
We report Giant  Metrewave Radio Telescope (GMRT) H{\sc i} 21\,cm imaging of CGCG~137$-$068, 
the host galaxy of the fast-evolving luminous transient (FELT) AT2018cow. This is the first 
study of the gas properties of a FELT host galaxy. We obtain a total H{\sc i} mass of 
$(6.6 \pm 0.9) \times 10^8$~M$_\odot$ for the host galaxy, 
which implies an atomic gas depletion time of 
$3$~Gyr and a gas-to-stellar mass ratio of $0.47$, consistent 
with values in normal star-forming dwarf galaxies. At spatial resolutions of $\geq 6$~kpc, the H{\sc i} of CGCG~137$-$068 appears to be distributed in a disk, in mostly regular rotation.
However, at spatial resolutions of $2$~kpc, the highest column density H{\sc i} is found to lie in an asymmetric ring around the central regions, with H{\sc i} column densities 
$\geq 10^{21}$~cm$^{-2}$; AT2018cow lies within this high column density ring. 
This H{\sc i} ring could be suggestive of an interaction between CGCG~137$-$068 and 
a companion galaxy.
Such a ring is ideal for the formation of compact regions of star formation hosting massive stars which are likely progenitors of FELTs.
We measure a 1.4 GHz flux density of $1.13 \pm 0.13$ mJy for AT2018cow on 2018 August 27.
\end{abstract}



\section{Introduction}
\label{sec:int}

Over the last decade, high-cadence wide-field optical surveys have resulted in the 
discovery of a new class of extragalactic variable sources, referred to as fast-evolving 
luminous transients \citep[``FELTs'';][]{drout14,arcavi16,tanaka16,rest18,pursiainen18}.
These transients 
show rapid rise and fading times, with, typically, $\lesssim 10$~days to peak luminosity 
and exponential decline $\lesssim 30$~days after the peak, and a wide range of peak 
luminosities, $-15 \geq M_g \geq -22.25$ \citep[e.g.][]{pursiainen18}. A black-body 
model provides a good fit to the optical spectral energy distribution upto a few weaks after the peak, yielding peak bolometric 
luminosities of $10^{42}$ to $10^{44}$~erg~s$^{-1}$ and peak temperatures of 
$8,000$ to $30,000$~K \citep[][]{drout14,pursiainen18}. 
The spectra are typically featureless blue continua, consistent with hot ejecta \citep{drout14,pursiainen18}. 

Nearly 100 FELTs have been detected till date, with spectroscopic redshifts available for 
about half of the host galaxies \citep[typically in the redshift range $0.2$ to $0.8$; e.g.][]{pursiainen18}. 
Although the wide range of FELT luminosities has been difficult 
to explain in theoretical models, \citet{pursiainen18} emphasize that all their events with 
identified hosts arise in star-forming galaxies. 
This favours models in which FELTs originate in short-lived massive progenitor stars \citep[see, e.g.,][]{kasen10,chevalier11,dexter13,kleiser14,rest18}.
Unfortunately, the rarity of FELTs has meant that most events have not been identified in real time, 
but rather in archival surveys, without the possibility of early multi-wavelength studies that 
might allow one to distinguish between the various models.

The remarkable optical transient AT2018cow was discovered on 2018 June 16 by the Asteroid 
Terrestrial-impact Last Alert System (ATLAS) survey \citep[][]{smartt18,prentice18}. With a rise time 
of $\sim 2.5$~days \citep{prentice18}, a blue spectrum, a black-body optical spectrum around 
the peak with a temperature of $30,000$~K, and a peak bolometric luminosity of 
$4 \times 10^{44}$~erg~s$^{-1}$ \citep{perley18}, it has characteristics similar 
to the FELTs discovered in earlier surveys. However, the early ATLAS identification of this 
transient allowed rapid follow-up observations over a wide range of frequencies, extending 
from the radio regime to X-rays and gamma rays \citep[e.g.][]{ho18,kuin18,perley18,margutti18}.
These studies have found evidence for a number of exciting new characteristics, including 
(1)~a very fast rise and fading, on timescales of a few days, (2)~an initially featureless optical 
spectrum, followed by a single broad absorption feature during days $4-8$ after the peak,
centred at $\sim 4600$\AA, followed by a weak, broad emission feature at $\approx 4850$\AA, and then a variety of similar emission 
lines, (3)~a high radio luminosity, (4)~long-lived millimetre-wave emission, extending to 
$\approx 670$~GHz, (5)~an additional source of X-ray emission from what appears to be a central 
engine, etc \citep[e.g.][]{prentice18,ho18,perley18,margutti18}. The detailed observational properties have 
proved challenging to explain so far, with some authors arguing for a supernova shock wave 
breaking out of a dense medium, others for a central engine providing extended energy injection, 
and yet others for a disruption of a stellar companion by an intermediate-mass black hole.
 
AT2018cow is located $\sim 5.9''$ \citep[equivalent to $1.8$~kpc,][]{perley18} from the centre of the dwarf spiral galaxy 
CGCG~137-068, at $z = 0.014145$ \citep{sdssdr6}. This is the first case of the discovery of a FELT 
in the nearby Universe, providing a unique opportunity to study physical conditions in the host 
galaxy of the transient, and the environment that gives rise to such an event.
Studies of the distribution and kinematics of atomic hydrogen through \hii emission line observations of FELT host galaxies are a promising tool to probe the environment that gives rise to the progenitors of these transients.
Such a study has provided important observational clues to the origin of a transient in the case of the nearest Gamma Ray Burst (GRB) host galaxy  \citep[][Arabsalmani et al., in press]{2015MNRAS.454L..51A}. 
In this {\it Letter}, we present a Giant Metrewave Radio Telescope (GMRT) \hii study of CGCG~137-068, the first case of \hii emission mapping of the neutral gas in a FELT host galaxy.

\section{Observations and data analysis}
\label{sec:obs}

We used the GMRT L-band receivers to observe CGCG~137-068 on 2018~August~27 in proposal DDTC022 
(PI: Arabsalmani). The observations used the GMRT Software Backend as the correlator, with a 
bandwidth of 16.67~MHz, centred at 1399.67~MHz and sub-divided into 512~channels, yielding 
a velocity resolution of $6.9$~\kms\ and a total velocity coverage of $3533$~\kms\
(at $z=0.014145$). Observations of the standard primary calibrators 3C286 and 3C48 at the start and
end of the run were used to calibrate the flux density scale, and of the nearby bright compact 
source 1609+266, to calibrate the antenna gains and passband shapes. The total on-source time was 
$\approx 5$~hours, with 27 working antennas.

The GMRT data were analysed in ``classic'' {\sc aips} \citep{greisen03}, following standard 
procedures. After initial data editing to remove dead antennas and data affected by radio frequency 
interference, and calibration of the complex gains and passband shapes, the calibrated 
visibilities were averaged together (excluding edge channels and $\sim 200$~\kms\ around the 
expected \hii redshifted line frequency of CGCG~137-068) to produce a multi-channel continuum 
data set with a frequency resolution of $\approx 1$~MHz (to avoid frequency-smearing in the 
continuum image). The final antenna-based gains were obtained via a standard iterative 
self-calibration and imaging procedure on this data set, with a few rounds of phase-only 
self-calibration and 3-D imaging, followed by amplitude-and-phase self-calibration, 3-D 
imaging, and further data editing. 
The imaging was done using the full uv-data set excluding baselines which are smaller than 1 kilo $\lambda$, with a uv-taper of 90 kilo $\lambda$ and robust weighting ($\lambda$ being the observation wavelength).
The average amplitude of the antenna gains was normalized 
during the amplitude-and-phase self-calibration, to preserve the flux density scale. This 
procedure was repeated until the image showed no improvement on further self-calibration. 
The entire GMRT primary beam was imaged, out to a radius of $\approx 0.3$~degrees, using 37 facets.

We subtracted out the final continuum image from the calibrated spectral-line visibilities using 
the task {\sc uvsub}, and next used the task {\sc cvel} to shift the residual visibilities to 
the heliocentric frame. The residual visibilities were then imaged with different tapering and 
weighting schemes, to produce spectral cubes at a range of angular resolutions. To increase the 
signal-to-noise ratio while deconvolving the array point spread function, the cubes were created 
by averaging either two or three channels, yielding final velocity resolutions of 
$13.9$~\kms\ or $20.9$~\kms.  Any residual continuum emission was removed by 
fitting a second-order polynomial in the image plane, using the task {\sc imlin}. The parameters 
of the different spectral cubes, made at different angular and velocity resolutions, are presented 
in Table~\ref{tab:cube}. Note that the cube with the coarsest angular resolution 
($43.1'' \times 39.1 ''$) and velocity resolution ($20.9$~km/s) was made in 
order to accurately estimate the total \hi mass of the galaxy, without resolving out any of the 
\hii emission.

\begin{table*}
\caption{\label{tab:cube}Parameters of the GMRT \hii data cubes. Column 1 lists the FWHM of the elliptical synthesized beam of each cube along their respective major and minor axes. Column 2 lists the maximum baseline length used, in units of observation wavelength, for making each cube. Column 3 lists the respective baseline for each cube, in observation wavelength units, at which the Gaussian function used to weight down long baseline data points is at 30\% of its peak. Column 4 gives the type of weighting applied to the uv-data when making each cube. Column 5 lists the channel widths of the cubes. Column 6 lists the RMS noise in the line-free channels of each cube.}
\begin{center}
\begin{tabular}{cccccc}
\hline
Synthesized Beam   &uv-maximum&uv-taper&weighting&Channel width & RMS noise\\
($'' \times ''$)   &(kilo $\lambda$)&(kilo $\lambda$)&&(\kms)&(mJy Bm$^{\rm -1}$)\\
\hline
$ 43.1\times 39.1$ &5&4.8&natural& $20.9$ & $0.7$ \\
$ 34.6\times 27.1$ &10&6&robust& $13.9$ & $1.0$ \\
$ 20.1\times 13.5$ &20&12&robust& $13.9$ & $0.8$ \\
$ 6.2\times 5.6$   &40&30&robust& $13.9$ & $0.6$ \\
\hline
\end{tabular}
\end{center}
\end{table*}

The task {\sc momnt} was then applied to the spectral cubes to obtain maps of the total \hii 
intensity and the intensity-weighted \hii velocity field at the chosen angular resolutions.
This task creates a secondary spectral cube by smoothing out the emission along both spatial 
and velocity axes, and then masks out pixels lying below a chosen threshold flux.  The mask 
is applied to the original spectral cube before creating the moment maps. The smoothing ensures 
that localized noise peaks are ignored, and that only emission that is correlated both spatially and 
along the velocity axis is selected. We smoothed our spectral cubes spatially with a Gaussian 
kernel of full width at half maximum (FWHM) equal to six pixels, and along the velocity axis 
by applying Hanning smoothing across blocks of three consecutive velocity channels. The threshold 
flux used to create the mask was typically $\approx 1.5$ times the root-mean-square noise in 
each channel of the original spectral cube.

The integrated \hii emission spectrum for CGCG 137$-$068 was derived from the lowest-resolution 
spectral cube, by summing over the flux within the spatial extent of the total \hii intensity 
(i.e. zeroth moment) image, in each channel of the cube.  The \hii spectrum and the moment images 
made from the higher-resolution data cubes are presented and discussed in the next section.

\section{Results}
\label{sec:results}

\begin{figure}
\begin{center}
\includegraphics[scale=0.42,trim={0.5cm 5.0cm 1.4cm 5.0cm},clip]{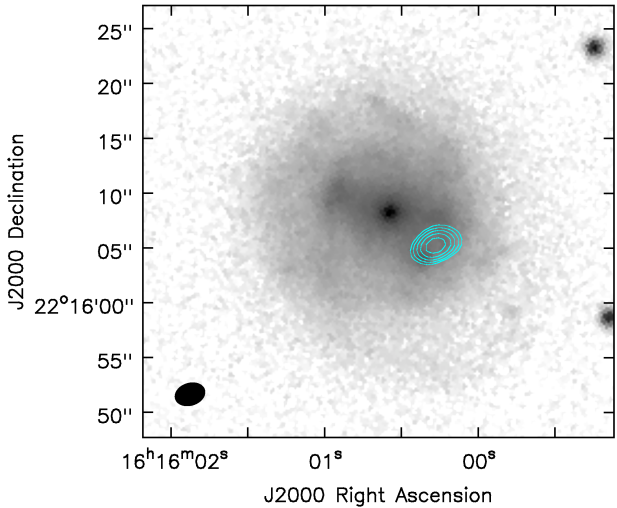}
\end{center}
\caption{The GMRT 1.4~GHz continuum image (in cyan contours) of CGCG~137$-$068, overlaid on 
an SDSS r-band image. The first 
contour is at $5\sigma$, where $\sigma$ is the RMS noise away from bright sources. Subsequent contours are at at intervals of $\sqrt{2} \times \sigma$. The GMRT synthesized beam 
is represented by the ellipse shown in the bottom left corner.}
\label{fig:cont}
\end{figure}

The GMRT 1.4~GHz continuum image, the central portion of which is shown in 
Fig.~\ref{fig:cont}, has an angular resolution of $2.8'' \times 2.1''$, and a 
root-mean-square (RMS) noise of $38 \; \mu$Jy/Bm, away from bright sources. An 
unresolved source is clearly visible at the centre of the image, at (J2000 co-ordinates) 
RA~=~16h16m00.3s, Dec.~=~22d16$'$05$''$, consistent within the errors with the position of 
AT2018cow \citep{bietenholz18}. We fit a 2-D Gaussian model to a small region centred on the 
source to obtain its flux density. 
Based on our earlier experience, systematic errors in the GMRT 
flux density scale are $\lesssim 10$\% at these frequencies, and with the above analysis 
procedure. 
The measured value for the 1.4~GHz flux density of AT2018cow on 2018~August~27 is 
$1.13 \pm 0.13$~mJy, where we have added the statistical error from the fit and the systematic error on the flux density scale in quadrature.

\begin{figure}
\begin{center}
\includegraphics[scale=0.375]{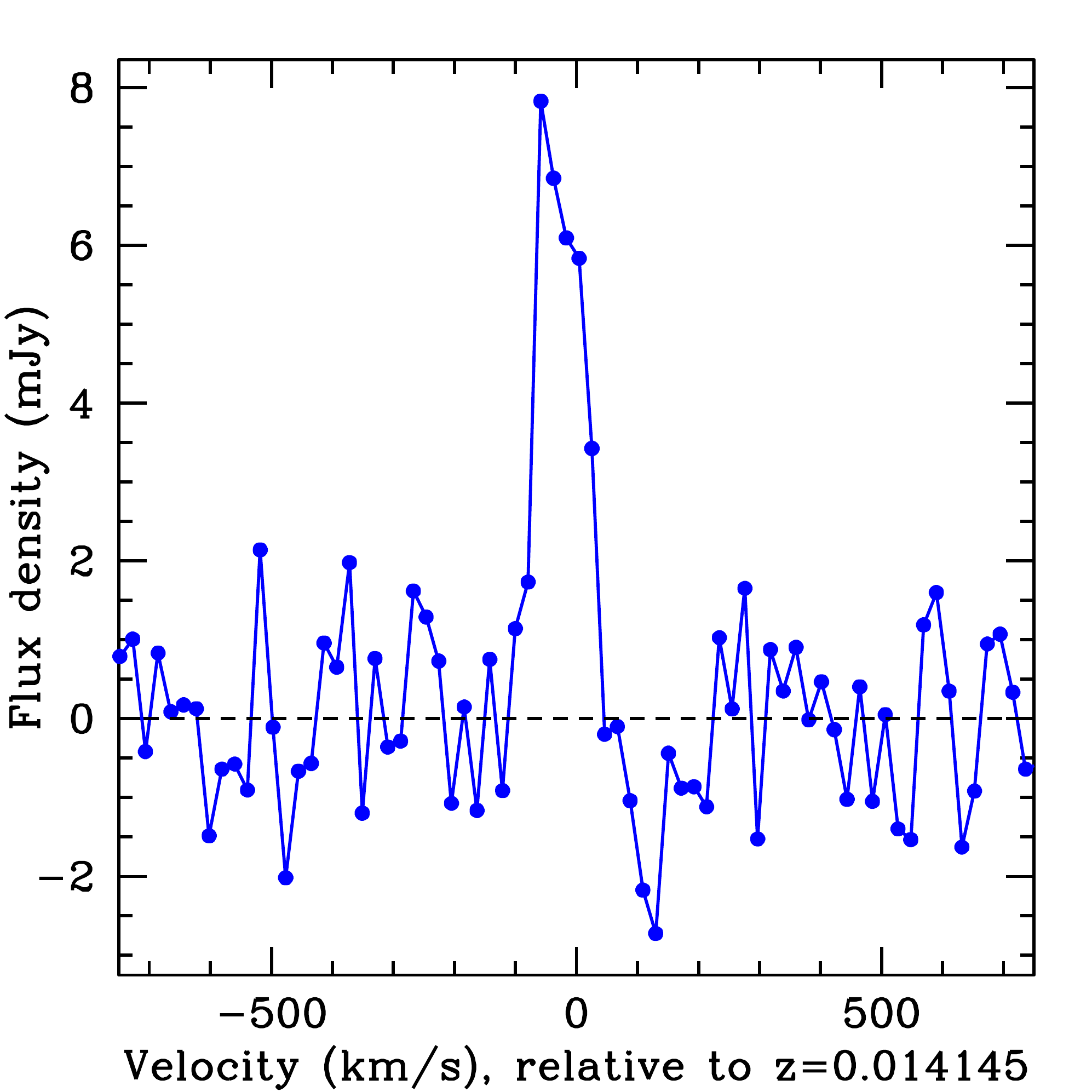}
\end{center}
\caption{The \hii emission spectrum for CGCG~137$-$068 derived from the lowest resolution 
spectral cube, as described in Sec.~\ref{sec:obs}.}
\label{fig:spec}
\end{figure}

The GMRT \hii emission spectrum of CGCG~137$-$068 is shown in Fig.~\ref{fig:spec}.
We measure an integrated \hii line flux density of $0.69 \pm 0.09$~Jy~\kms, where the quoted error combines, 
in quadrature, the statistical error and an assumed 10\% systematic error on the flux scale. 
Using a luminosity distance of $63.6$~Mpc \footnote{We assume a flat Lambda Cold Dark 
Matter cosmology, with $H_0 = 67.4$~\kms~Mpc$^{-1}$, $\Omega_m = 0.315$, and 
$\Omega_\Lambda = 0.685$ \citep{planck16}.}, we obtain a total \hi mass of 
$(6.6 \pm 0.9) \times 10^8 {\rm M}_{\odot}$ for CGCG~137$-$068.
The \hii emission profile has a velocity width of $W_{50} \approx 93$ \kms.

\begin{figure*}
\begin{center}
\begin{tabular}{ccc}
\includegraphics[scale=0.3,trim={1.0cm 5.0cm 1.4cm 5.0cm},clip]{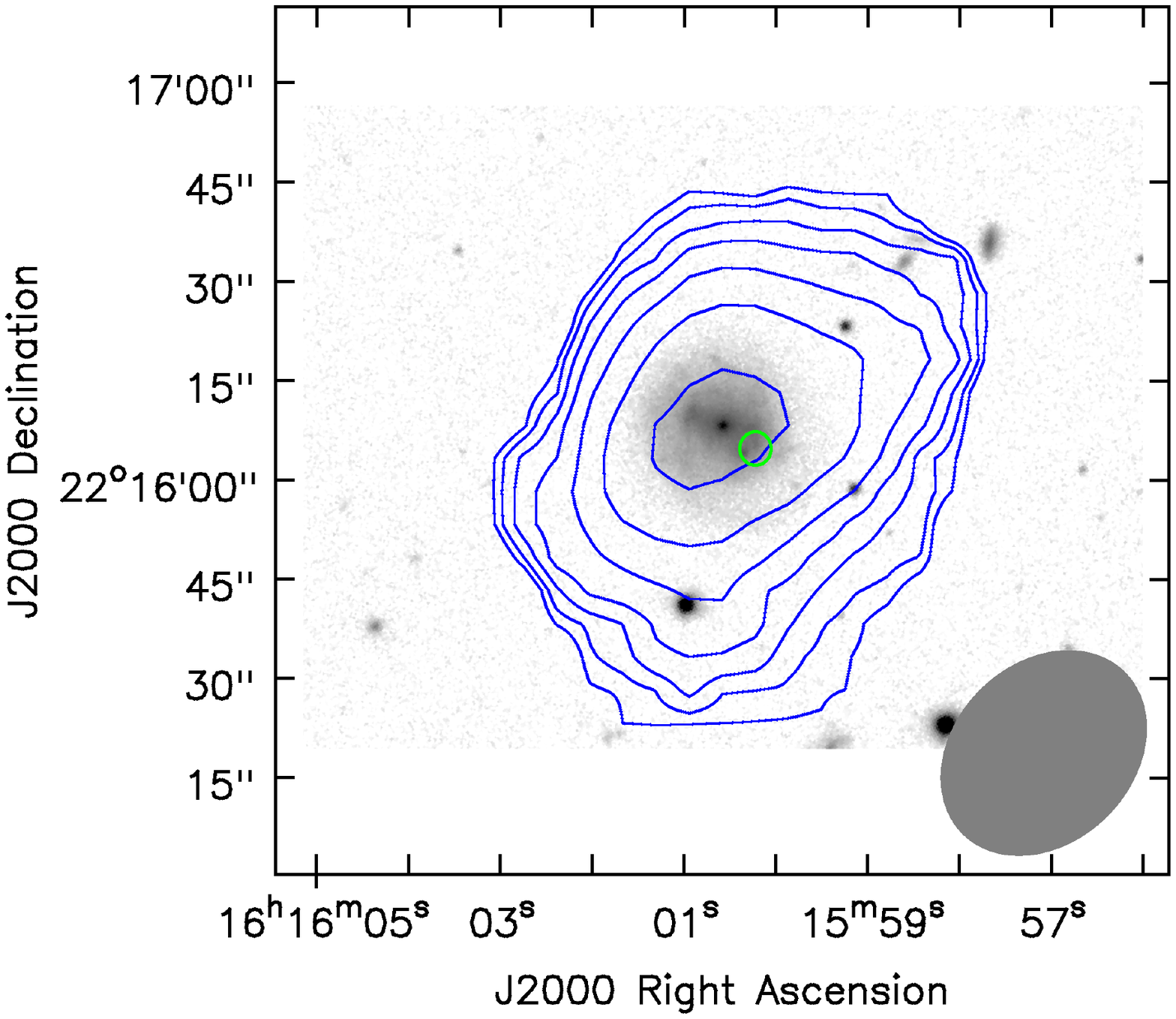} & 
\includegraphics[scale=0.3,trim={1.2cm 5.0cm 1.4cm 5.0cm},clip]{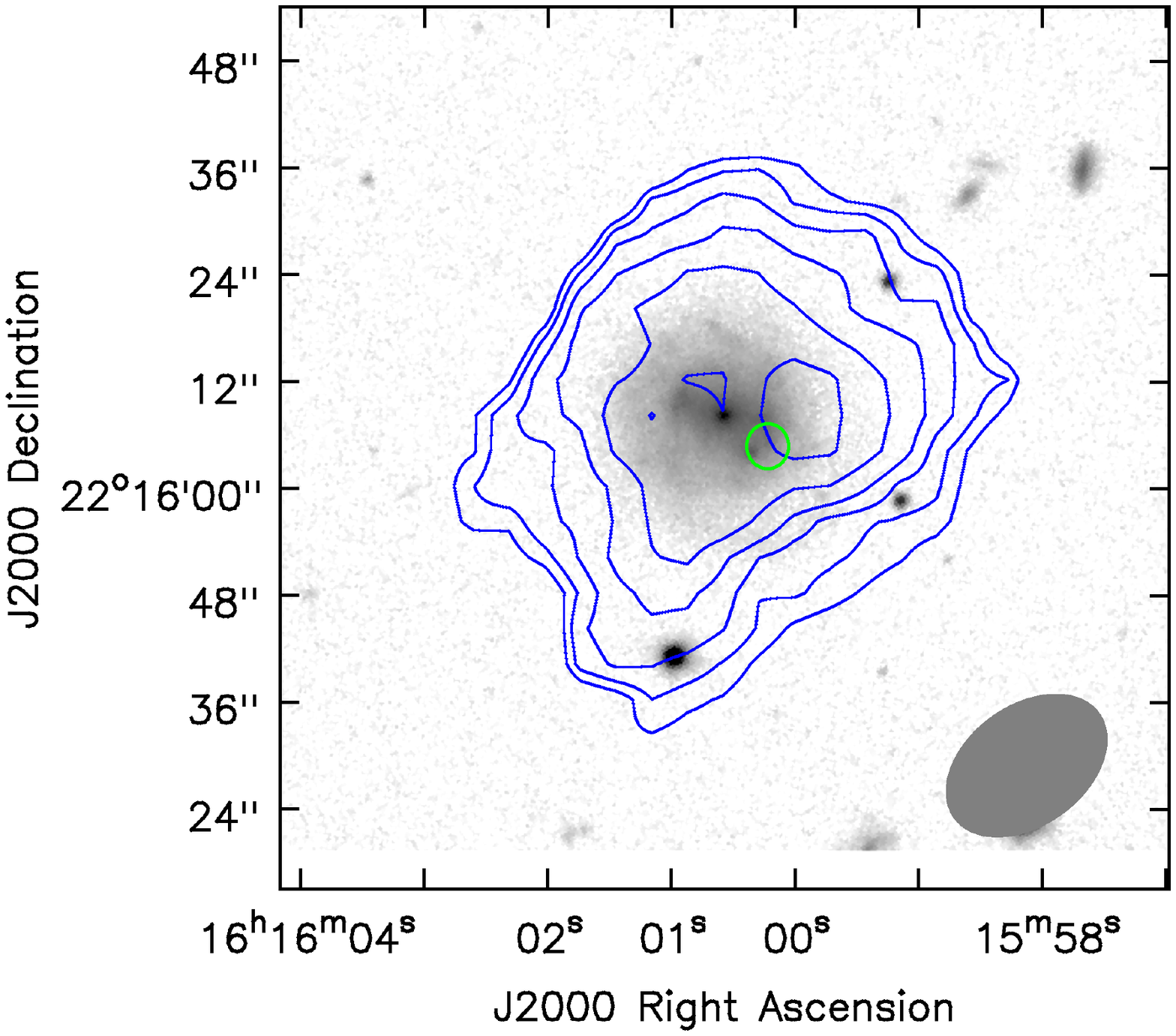} & 
\includegraphics[scale=0.3,trim={1.2cm 5.0cm 1.4cm 5.0cm},clip]{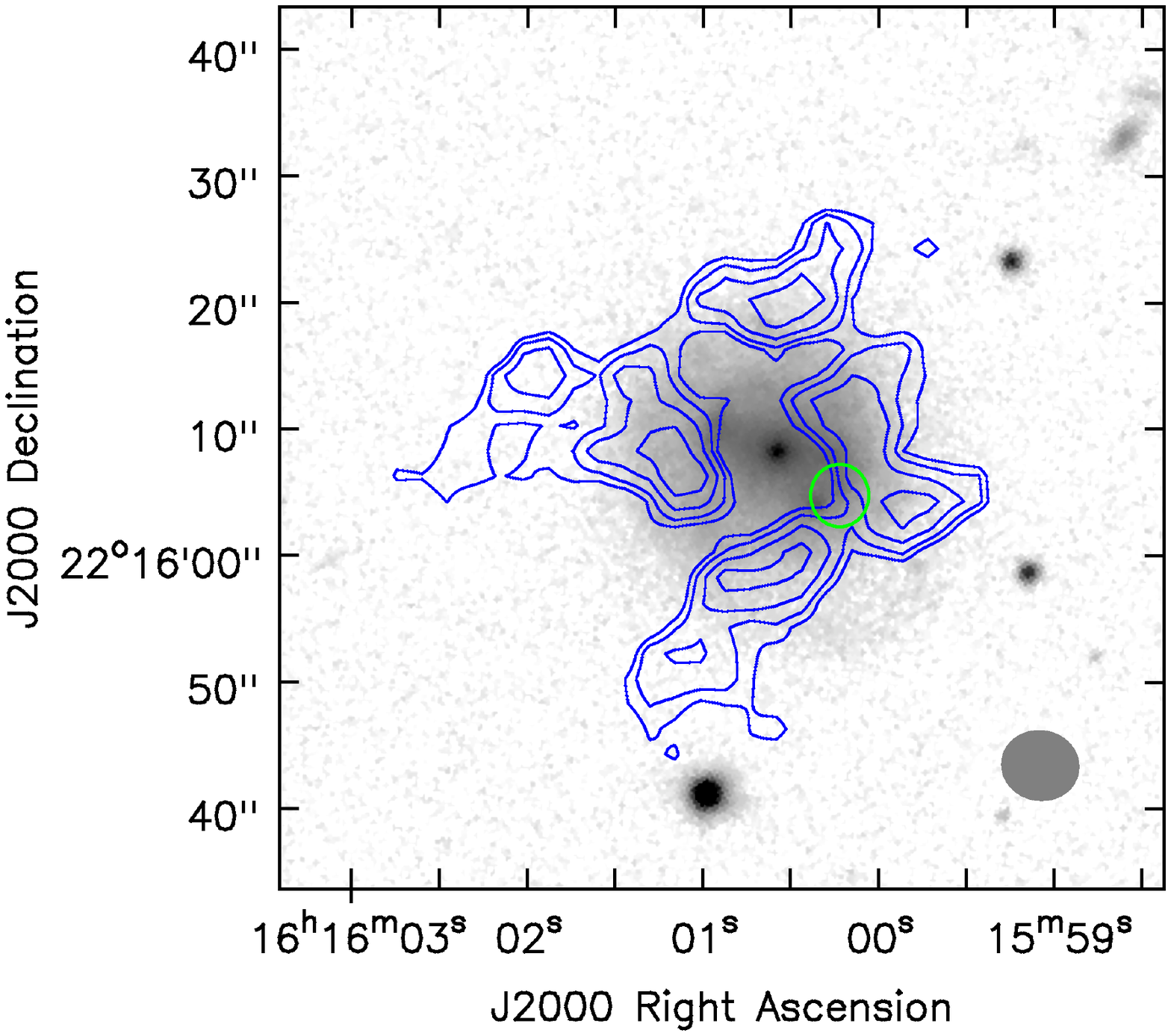} \\
\includegraphics[scale=0.3,trim={1.0cm 5.0cm 1.4cm 5.0cm},clip]{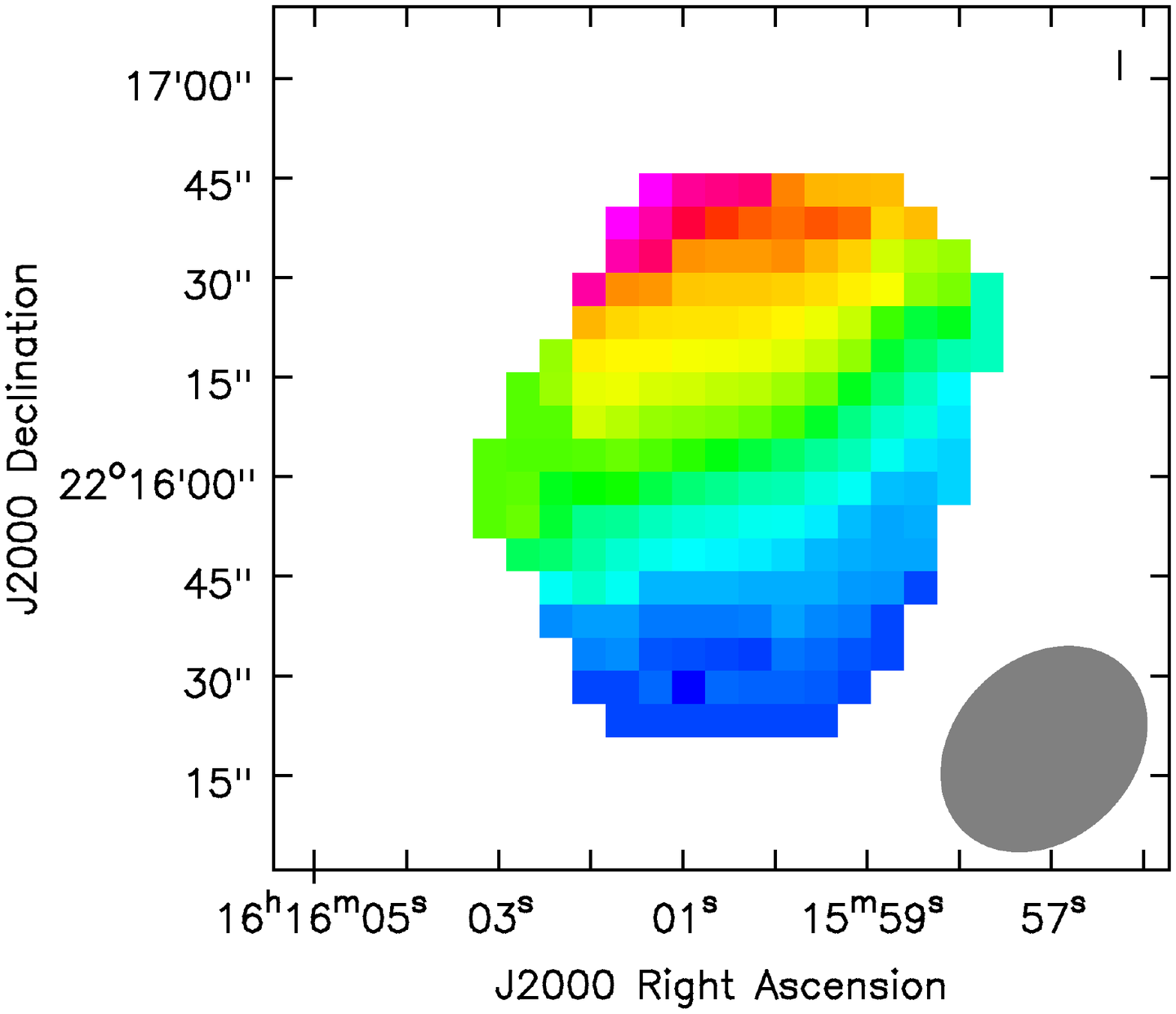} & 
\includegraphics[scale=0.3,trim={1.2cm 5.0cm 1.4cm 5.0cm},clip]{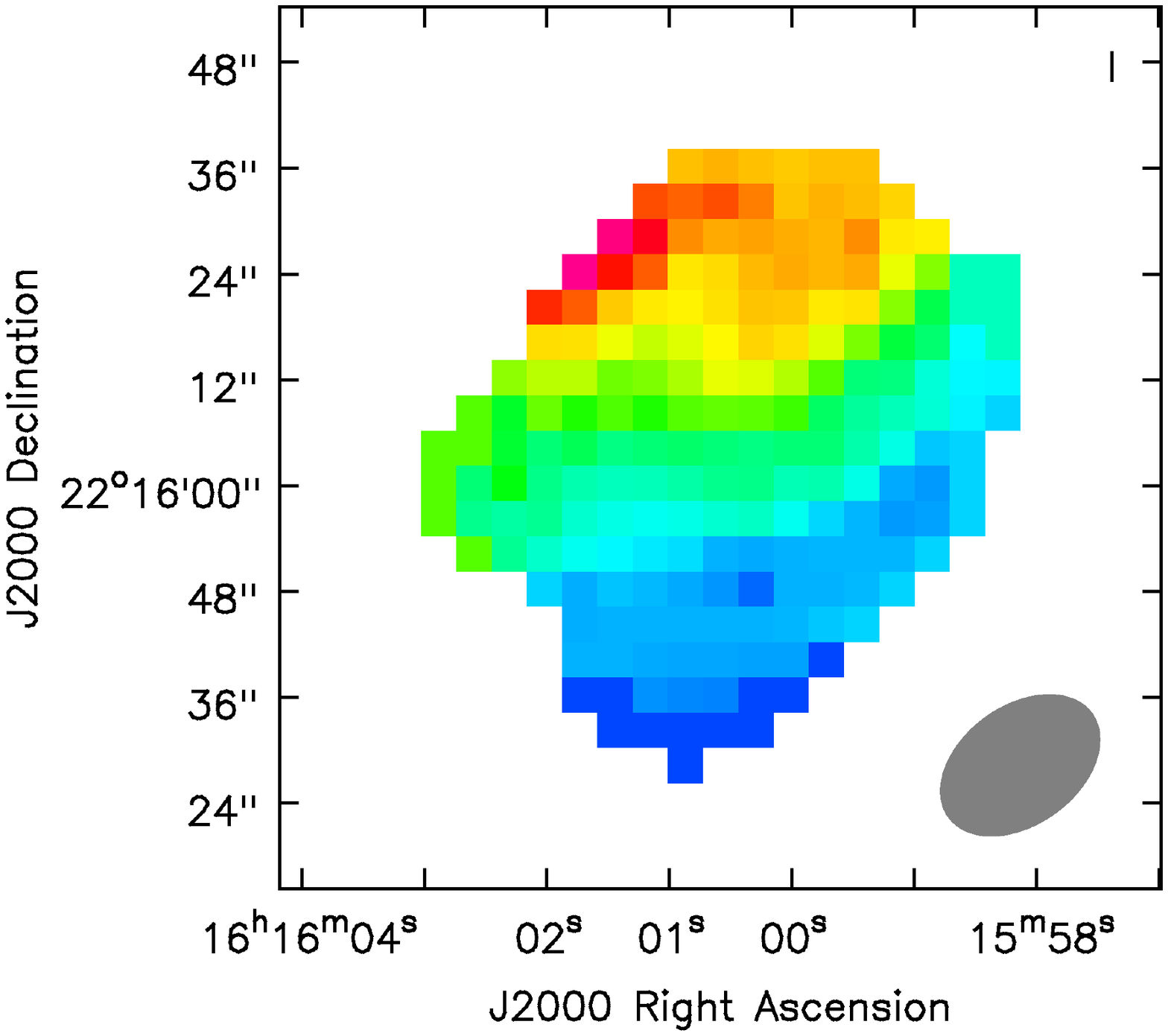} & 
\includegraphics[scale=0.3,trim={1.2cm 5.0cm 1.4cm 5.0cm},clip]{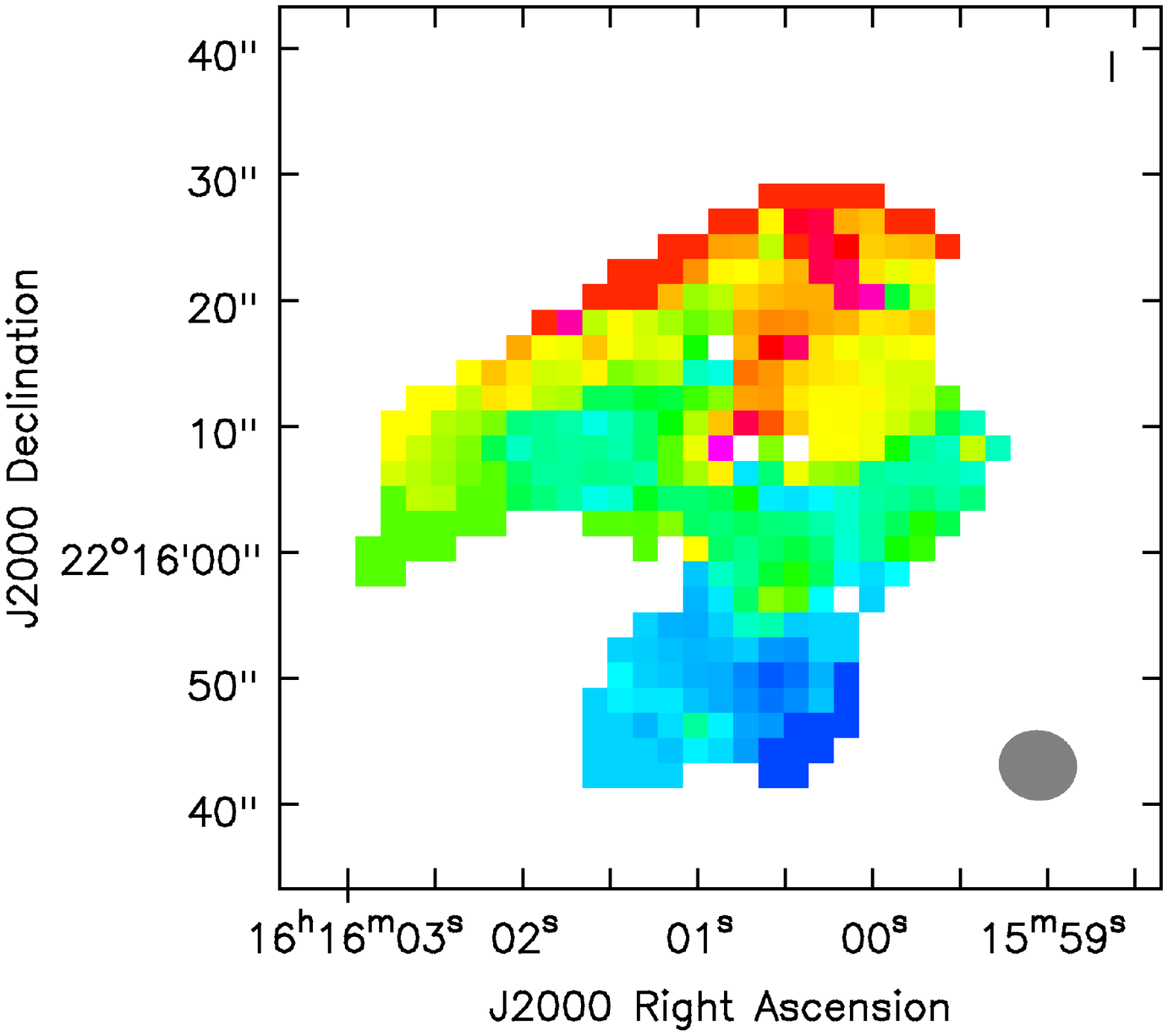} \\
\end{tabular}
\end{center}
\includegraphics[scale=0.6]{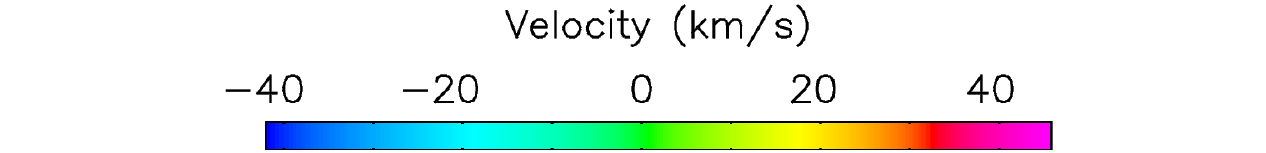} 
\caption{{\it Top row:} Integrated \hii intensity (contours) overlaid on an SDSS r-band image of 
CGCG 137$-$068 (greyscale) at different spatial resolutions: 34.6\arcs$\times$27.1\arcs\
(left panel), 20.1\arcs$\times$13.5\arcs\ (middle panel), and 6.2\arcs$\times$5.6\arcs. The beam 
sizes are shown in the bottom-right corner of each panel. Note that the field-of-view also changes and becomes tighter as we move from left to right panels.  The first 
contours of the three \hii intensity maps are at $5.3 \times 10^{19}$~\cm\ (left panel), 
$1.4 \times 10^{20}$~\cm\ (middle panel), and $7.7 \times 10^{20}$~\cm\ (right panel), with the 
\hii intensity at each subsequent contour in each map increasing by a factor of $\sqrt{2}$. 
The green circle in each panel marks the location of the transient AT2018cow, and is 5\arcs in diameter.
{\it Bottom row:} The \hii velocity field at the spatial resolutions of the panels 
immediately above, with the beam size again in the bottom-right corner of each panel. The 
velocity colour bar is shown below the three panels, relative to $z = 0.014145$. }
\label{fig:ov}
\end{figure*}

We study the spatial structure and kinematics of the \hi in CGCG~137$-$068 using the zeroth 
and first velocity moment maps of the \hii emission derived from spectral cubes created
at three different resolutions (the three highest resolutions listed in Table~\ref{tab:cube}). 
The moment maps are presented in Fig.~\ref{fig:ov}, where the angular resolutions of the images 
are $34.6'' \times 27.1''$ (left panel), $20.1'' \times 13.5''$ (middle panel), and 
$6.2'' \times 5.6''$ (right panel).
Note that as we improve the angular resolution of the \hii images, we are less sensitive to low column densities.
Thus, making the images at different resolutions helps us capture both the extent of the diffuse atomic gas and the structure of the atomic gas on smaller scales.
The \hii column density (contours) overlaid on a Sloan Digital Sky Survey \citep[SDSS;][]{sdssdr6} Data Release~6 r-band image (in greyscale) are shown in the panels of the top row in Fig.~\ref{fig:ov}, while those of the bottom row show the \hii velocity field. It is clear from the left and middle panels
(spatial resolutions $\geq 6$ kpc) 
that the \hi in CGCG~137$-$068 is distributed in a disk which extends beyond the optical disk of the 
galaxy. The velocity fields in the left and middle panels indicate that the \hi is in regular rotation.

Our highest-resolution \hii images of CGCG~137$-$068 (spatial resolution $\approx 2$ kpc) are shown 
in the right panels of 
Fig.~\ref{fig:ov}. 
At this angular resolution, the image is only sensitive to gas at a high \hi column density, $\geq 10^{21}$~cm$^{-2}$. The high column density 
atomic gas is seen to lie in an asymmetric ring around the optical centre of the galaxy. The transient 
AT2018cow (location indicated by the green circle in the top panels) lies within this ridge of high 
column density gas; the \hii column density around the location of AT2018cow averaged over the area of the beam is $1.1 \times 10^{21}$~\cm.

The velocity structure of the high column density gas at this high angular resolution is particularly interesting. The rotation evident in the lower resolution maps is much more broken and 
non-uniform at this high resolution.
We note that there exist large velocity gradients within the ring, as high as $\sim 40$ \kms\ over $2$ kpc scales.

\section{Discussion}

CGCG~137$-$068 has a stellar mass of $\rm 10^{9.15^{+0.05}_{-0.10}}\; M_{\odot}$ and a star 
formation rate (SFR) of $\rm 0.22 \; M_{\odot}\,yr^{-1}$ \citep{perley18}. 
These properties are consistent with a ``main sequence'' galaxy in the M$_*$--SFR plane 
in the nearby Universe \citep[][]{Brinchmann04-2004MNRAS.351.1151B}. Our H{\sc i} mass 
estimate of $6.6 \times 10^8$~M$_\odot$ then yields a gas-to-stellar mass ratio of $0.47$, 
consistent with values in normal star-forming galaxies with similar stellar masses 
\citep[e.g.][]{Denes14-2014MNRAS.444..667D}. Comparing the \hi mass with the SFR yields an atomic gas depletion 
time of $3$~Gyr, again consistent with typical values in star-forming galaxies in the 
nearby Universe \citep[see, e.g., the xCOLDGASS sample;][]{saintonge17}.
By comparing the width of the \hii emission of CGCG~137$-$068 with its combined gas$+$stellar mass, we find that the galaxy follows the baryonic Tully-Fisher relation \citep{zaritsky14}.
 Overall, 
both the global stellar and \hi properties of CGCG~137$-$068 are consistent with its being a
normal star-forming dwarf galaxy, on the main sequence.

While the global stellar and gas properties of CGCG~137$-$068 are typical of normal 
 star-forming galaxies, the same cannot be said of the spatial distribution or the velocity field 
of the atomic gas, as seen in the highest-resolution GMRT \hii images of Fig.~\ref{fig:ov}.
The highest column density \hi appears to be arranged in a asymmetric ring around the central 
regions of the galaxy. Such a high-column density ring of atomic gas 
is very unusual for a low-mass dwarf galaxy such as CGCG 137$-$068.
The transient AT2018cow is located within this ring of high column density \hi gas.
This suggests that the occurrence of AT2018cow might be related to the compression
of gas to high densities within the ring, that could have caused a rapid burst of star formation.

Similar evidence that a highly energetic event took place in the highest column density \hi in a galaxy
was found by \citet{2015MNRAS.454L..51A} based on GMRT \hii mapping observations of the 
host galaxy of GRB 980425. There too, the high column density 
\hi appears to be in a ring, which is the site of actively star-forming regions in the galaxy.
Deeper follow-up \hii observations, combined with archival Very Large Telescope and Spitzer Space Telescope 
imaging data, have revealed the presence of a faint companion galaxy interacting with the
GRB host galaxy (Arabsalmani et al., in press).
Numerical simulations of galaxy-galaxy mergers by Arabsalmani et al. (inn press) show that a 
collisional interaction between 
the GRB host galaxy and the companion can explain the formation of the high column 
density H{\sc i} ring. Their simulations indicate that a collisional gas ring should 
have velocity gradients of a few tens of \kms\ on sub-kpc scales within the ring. 

Regions of compact and intense star formation like super star-clusters can produce a number of 
high-mass stars which are the progenitors of highly energetic stellar events like GRBs, and the 
likely progenitors of FELTs. Interacting galaxies are ideal sites for the formation of such 
super star-clusters, due to the formation of compact and massive Giant Molecular Clouds (GMCs) in such 
systems \citep[e.g.][]{jog92,barnes96,mihos96,renaud09}. Massive GMCs form efficiently in 
interacting galaxies due to the absence of gravitational shear and the increased turbulence, 
which both aid in the collapse of large amounts of gas.
The large velocity gradients within the interstellar gas of such systems also increases the Jeans mass of collapsing clouds. 
The increased Jeans mass increases the temperature of the clouds, thereby shifting the stellar mass function towards 
the high mass end \citep{Elmegreen-1993ApJ...412...90E}.  
Massive super star-clusters (SSCs) have indeed been found more commonly in interacting systems 
\citep[e.g.][]{Elmegreen-1993ApJ...412...90E,deGrijs03-2003NewA....8..155D,Bastian08-2008MNRAS.390..759B}.
In particular, collisional interactions have been shown to be efficient in forming fewer but more massive and compact 
SSCs compared to tidal interactions \citep[see][and references therein]{Renaud18-2018MNRAS.473..585R}. 
Therefore, a connection between interacting galaxies and highly energetic transients born out of massive stars appears to be a plausible one.

After GRB~980425, AT2018cow is the second case of a luminous transient associated with an asymmetric ring 
of high column density atomic gas in the host galaxy. 
The velocity gradient in the western section of the ring 
(which contains AT2018cow) reaches $\sim 40$~\kms\ on the spatial scales of $2$ kpc (our highest resolution), similar to what is seen for the \hi ring in the host galaxy of GRB 980425.
The observed velocity gradient is relatively large for a low-mass galaxy like CGCG~137$-$068.
Deeper and higher resolution \hi observations are required to confirm the existence of high
velocity gradients on sub-kpc scales within the ring of high column density \hi.

We note, in passing, that the William Herschel Telescope r-band optical image of CGCG 137-068 \citep{perley18} shows indications of a stellar ring, roughly coincident with the HI ring reported here.
Deep multi-wavelength observations are required to confirm whether GMCs and regions of dense
star formation are located along the \hi ring, especially around the location of AT2018cow.

In summary, we have used the GMRT to carry out \hii spectroscopy of CGCG~137$-$068, the host 
galaxy of the fast-evolving luminous transient AT2018cow.
This is the first study of the gas properties 
of a FELT host galaxy. Our $\geq 6$~kpc spatial resolution GMRT \hii maps indicate 
that CGCG~137$-$068 is a disk galaxy, with the \hi mostly in regular rotation, and with an \hi mass 
of $\rm M_{HI} = (6.6 \pm 0.9) \times 10^8$~M$_\odot$. 
We obtain a gas-to-stellar mass ratio of $0.47$ and a gas depletion time of 
$3$~Gyr, both consistent with typical values found in normal star-forming dwarf 
galaxies in the nearby Universe. However, our $2$~kpc spatial resolution \hii image 
shows that the high column density atomic gas is distributed in an asymmetric ring around the central regions of the galaxy and with high velocity gradients.
AT2018cow is located within this high \hi column density ring. 
Such a ring is an ideal site for the formation of super star-clusters hosting massive stars which are the likely progenitors of luminous transients like AT2018cow.


\section*{Acknowledgments}
MA would like to thank Emeric Le Floc'h and Frederic Bournaud for helpful discussions.
MA acknowledges support from  UnivEarthS Labex program at Sorbonne Paris Cit\'e (ANR-10-LABX-0023 and ANR-11-IDEX-0005-02). 
NK acknowledges support from the Department of Science and Technology via a Swarnajayanti Fellowship 
(DST/SJF/PSA-01/2012-13). We thank the staff of the GMRT who have made these observations possible. 
The GMRT is run by the National Centre for Radio Astrophysics of the Tata Institute of 
Fundamental Research. 

\bibliographystyle{mnras}
\bibliography{atcow}

\bsp
\label{lastpage}
\end{document}